\documentclass[acmsmall,screen,nonacm]{acmart}
\usepackage{booktabs}
\usepackage{wrapfig}
\usepackage{listings}
\usepackage{tikz}
\usetikzlibrary{positioning,arrows.meta,shapes.geometric,fit,calc}
\usepackage{xcolor}
\usepackage{multirow}
\usepackage{enumitem}
\usepackage{xspace}
\usepackage{pifont}
\setlist{nosep,leftmargin=*}

\newcommand{\sys}{\textsc{Quelle}\xspace}

\newcommand{\sem}[1]{\llbracket #1 \rrbracket}
\newcommand{\eqv}{\simeq}
\newcommand{\U}{\mathcal{U}}

\tikzset{qsym/.style={baseline={([yshift=-0.6ex]current bounding box.center)}},
  qw/.style={line width=0.4pt}, qg/.style={draw,fill=white,line width=0.4pt,inner sep=1pt,
  minimum size=7pt,font=\tiny\sffamily}}
\newcommand{\qbox}[1]{\tikz[qsym]{\draw[qw] (-7pt,0)--(7pt,0); \node[qg] at (0,0) {#1};}}
\newcommand{\qbbox}[1]{\tikz[qsym]{\draw[qw] (-8pt,0)--(8pt,0) (-8pt,-5pt)--(8pt,-5pt);
  \node[qg,minimum height=9pt,minimum width=10pt] at (0,-2.5pt) {#1};}}
\newcommand{\qcx}{\tikz[qsym]{\draw[qw] (-7pt,0)--(7pt,0) (-7pt,-5pt)--(7pt,-5pt) (0,0)--(0,-7.2pt);
  \fill (0,0) circle (1.2pt); \draw[qw] (0,-5pt) circle (2.2pt);}}
\newcommand{\qccx}{\tikz[qsym]{\draw[qw] (-7pt,0)--(7pt,0) (-7pt,-4.5pt)--(7pt,-4.5pt) (-7pt,-9pt)--(7pt,-9pt)
  (0,0)--(0,-11.2pt); \fill (0,0) circle (1.2pt) (0,-4.5pt) circle (1.2pt); \draw[qw] (0,-9pt) circle (2.2pt);}}
\newcommand{\qcz}{\tikz[qsym]{\draw[qw] (-7pt,0)--(7pt,0) (-7pt,-5pt)--(7pt,-5pt) (0,0)--(0,-5pt);
  \fill (0,0) circle (1.2pt) (0,-5pt) circle (1.2pt);}}
\newcommand{\qswap}{\tikz[qsym]{\draw[qw] (-7pt,0)--(7pt,0) (-7pt,-5pt)--(7pt,-5pt) (0,0)--(0,-5pt)
  (-1.8pt,-1.8pt)--(1.8pt,1.8pt) (-1.8pt,1.8pt)--(1.8pt,-1.8pt) (-1.8pt,-6.8pt)--(1.8pt,-3.2pt) (-1.8pt,-3.2pt)--(1.8pt,-6.8pt);}}

\lstdefinelanguage{qasm}{
  morekeywords={OPENQASM,include,qreg,creg,gate,measure,barrier,rz,sx,x,cx,h,cp,swap},
  sensitive=true, morecomment=[l]{//}}
\definecolor{lstkw}{HTML}{3A72B8}
\definecolor{lstcall}{HTML}{0F7A6C}
\definecolor{lstbuiltin}{HTML}{B85A28}
\definecolor{lstcom}{HTML}{8A8F98}
\definecolor{lstbg}{HTML}{F5F6F8}
\lstdefinestyle{py}{language=Python, basicstyle=\ttfamily\footnotesize, keywordstyle=\color{lstkw}\bfseries,
  commentstyle=\itshape\color{lstcom}, stringstyle=\color{lstbuiltin},
  emph={[1]h,p,cp,cu,ry,cx,cz,ccx,mcp,compose,measure}, emphstyle={[1]\color{lstcall}},
  emph={[2]range,enumerate,synth_qft_full,QuantumCircuit,pi}, emphstyle={[2]\color{lstbuiltin}},
  backgroundcolor=\color{lstbg}, framesep=3pt, frame=single, rulecolor=\color{lstbg},
  xleftmargin=3pt, xrightmargin=3pt, breakindent=2em, postbreak=\mbox{$\hookrightarrow$\ },
  numbers=none, showstringspaces=false}

\graphicspath{{figures/}}
\begin{document}

\title{Decompiling Quantum Assembly into Structured Programs}
\author{Pingchuan Ma}
\orcid{0000-0001-7680-2817}
\affiliation{
  \institution{Zhejiang University of Technology}
  \city{Hangzhou}
  \country{China}
}
\email{pma@zjut.edu.cn}

\author{Zhantong Xue}
\orcid{0009-0000-3419-9431}
\affiliation{
  \institution{The Hong Kong University of Science and Technology}
  \city{Hong Kong}
  \country{China}
}
\email{zxueai@cse.ust.hk}

\author{Zongjie Li}
\orcid{0000-0002-9897-4086}
\affiliation{
  \institution{The Hong Kong University of Science and Technology}
  \city{Hong Kong}
  \country{China}
}
\email{zligo@cse.ust.hk}

\author{Qixin Zhang}
\affiliation{
  \institution{Nanyang Technological University}
  \city{Singapore}
  \country{Singapore}
}
\email{qixin.zhang2026@gmail.com}

\author{Yuxuan Du}
\orcid{0000-0002-1193-9756}
\affiliation{
  \institution{Nanyang Technological University}
  \city{Singapore}
  \country{Singapore}
}
\email{yuxuan.du@ntu.edu.sg}

\author{Zhaoyu Wang}
\orcid{0009-0009-6892-1264}
\affiliation{
  \institution{The Hong Kong University of Science and Technology}
  \city{Hong Kong}
  \country{China}
}
\email{zwangjz@cse.ust.hk}

\author{Yuguang Zhou}
\orcid{0000-0001-5069-4563}
\affiliation{
  \institution{The Hong Kong University of Science and Technology}
  \city{Hong Kong}
  \country{China}
}
\email{yzhougv@cse.ust.hk}

\author{Shuai Wang}
\orcid{0000-0002-0866-0308}
\affiliation{
  \institution{The Hong Kong University of Science and Technology}
  \city{Hong Kong}
  \country{China}
}
\email{shuaiw@cse.ust.hk}

\author{Xiaoqin Zhang}
\orcid{0000-0003-0958-7285}
\affiliation{
  \institution{Zhejiang University of Technology}
  \city{Hangzhou}
  \country{China}
}
\email{xqzhang@zjut.edu.cn}

\renewcommand{\shortauthors}{Ma et al.}

\begin{abstract}
Quantum compilers translate programs into native gates, insert \texttt{SWAP} gates to route interactions onto a device, and optimize the result. The output is quantum assembly, a flat gate list that hides the algorithm's structure (QFT, Grover iteration, QAOA layers).
Understanding, auditing, porting, or reusing such assembly requires recovering a program that exposes this structure.

We present \sys, a decompiler that recovers structured programs (library calls, loops, functions, symbolic angles) from routed and optimized quantum assembly (OpenQASM) and checks every output for equivalence with its input.
\sys is built on one invariant: every step is either an \emph{exact rewrite}, checked where it is applied, or a \emph{hypothesis}, checked against the input before use.
It first lifts the assembly into a circuit by removing compilation artifacts, notably \emph{un-routing}, which recovers the \texttt{SWAP}s that routing inserted, including those merged into other gates. It then recovers algorithm templates (\emph{sketches}) whose parameters are \emph{solved} from quantities preserved by compilation, and, for algorithms without a template, library calls and loops by anti-unification.
Failed hypotheses are rejected; the corresponding operations remain in gate-level form.

On 816 assembly programs from 22 algorithm families and a random-circuit control, compiled at four routing and optimization levels with native two-qubit gates CX, ECR, or CZ, \sys's output is equivalent to the input for 100\% of the inputs, and it fully decompiles 94\% of them; compared on the same inputs, it fully decompiles 1.7$\times$ as many as the best LLM-based decompiler, which spends 15K tokens per input.
On nine algorithm families for which it has no template, the structure it recovers without templates explains 60\% of the gates.
It checks inputs of up to 2.3 million gates and emits no inequivalent output on 237 real-world QASMBench assembly files.

\end{abstract}


\maketitle

\section{Introduction}
\label{sec:intro}
\begin{figure}[t]
\centering
\includegraphics[width=\textwidth]{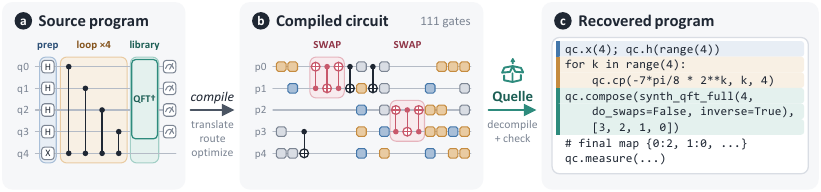}
\caption{Running example of \sys.}
\label{fig:motivation}
\end{figure}
Quantum programs are written at a high level of abstraction (a quantum Fourier transform, a Grover iteration, a QAOA layer), but they are \emph{executed} at a very different one.
A quantum device runs a \emph{circuit}, a straight-line program whose instructions (\emph{gates}) are small unitary matrices applied to one or two qubits at a time (\S\ref{sec:primer} summarizes the quantum notions we use). Written out as text, a circuit is a \emph{quantum assembly} program, usually in OpenQASM~\cite{openqasm2,openqasm3}, the Open Quantum Assembly Language.
The quantum compiler that produces this assembly works much like a classical back end~\cite{qiskit,tket,sabre}.
Instruction selection lowers every gate to the few native gates the device supports.
Qubit mapping, the analogue of register allocation, assigns the program's logical qubits to physical ones and inserts \texttt{SWAP} gates that move qubits around, because a two-qubit gate can act only on qubits that are adjacent on the chip.
Peephole and resynthesis optimizations then merge and cancel gates across the boundaries of the program's building blocks.
The result is a flat assembly program of native gates whose size grows with the routing overhead and from which the source-level structure is gone.
A 5-qubit phase estimation program becomes 111 assembly instructions over $\{\texttt{rz},\texttt{sx},\texttt{x},\texttt{cx}\}$, 27 of which are routing artifacts (Figure~\ref{fig:motivation}).

Quantum assembly nevertheless circulates widely. Since 2024, IBM's cloud service
has executed only circuits that are already compiled to the target device's
native gates and connectivity~\cite{ibmisa}, so every circuit submitted to it is
assembly; and benchmark suites distribute compiled circuits: 120 of the 252
QASMBench~\cite{qasmbench} files are transpiled, and MQT Bench~\cite{mqtbench}
also provides its benchmarks mapped to devices. Working with such an artifact
(to understand what it computes, to audit it, to port it to another device, to
compare it to a reference implementation, or to reuse it as a subroutine)
requires \emph{decompilation}: recovering a high-level program that is
equivalent to the assembly program and exposes its algorithmic structure.
Decompilation is well studied for classical binaries~\cite{cifuentes,phoenix}
and has recently reached deep-learning executables~\cite{btd}, and a growing
body of work now evaluates the semantic consistency and fidelity of decompiler
output~\cite{cao2024evaluating,dramko2024taxonomy}. For quantum circuits,
DeQompile~\cite{deqompile} evolves a program that generates a family of circuits
by genetic programming, guided by syntactic similarity and without checking
equivalence; its authors report lower scores on transpiled circuits. Existing
quantum tooling moves in the opposite direction, from programs to assembly
(transpilation, optimization, and verified
compilation~\cite{voqc,certiq,giallar}); circuit \emph{simplifiers} such as
ZX-calculus reduction~\cite{pyzx} shrink a circuit but do not recover a program
from it.

\paragraph{Key observation.}
Compilation changes the \emph{syntax} of a program but preserves its \emph{semantics}.
Different compilers, optimization levels, and chip layouts produce different gate sequences for the same algorithm, yet the matrix each region computes, and the invariants that identify the algorithm inside it, are unchanged.
Routing only relabels wires, so a routed circuit is the logical circuit up to a permutation, wherever its \texttt{SWAP}s have been merged.
A routed and optimized QAOA circuit is still a sequence of rotations whose angles are the algorithm's parameters, followed by a fixed change of basis (a Clifford frame, \S\ref{sec:compile}).
An optimized phase-estimation circuit is still a matrix whose diagonal blocks are successive powers of one operator.
This has two consequences.
First, the artifacts of compilation can be removed by \emph{exact} semantic rewrites, so lifting can be sound by construction.
Second, once the artifacts are removed, the parameters of an algorithm template (the angles of a QAOA layer, the eigenphase of a QPE, the marked state of a Grover search) can be \emph{solved for} from the circuit, and the resulting hypothesis can be \emph{checked} with exact equivalence oracles.

\paragraph{Challenges.}
Turning this observation into a decompiler raises four challenges.
\textbf{(C1) Hidden routing.} A \texttt{SWAP} inserted by the router is frequently absorbed into a neighboring two-qubit gate. A generic two-qubit gate already needs three CNOTs, and so does the same gate merged with a \texttt{SWAP}, so neither pattern matching nor gate counts reveal it.
\textbf{(C2) Approximate recovery vs.\ exact guarantee.} Recovered parameters are real numbers obtained from numerical linear algebra, yet the output must be equivalent to the input, and be checked to be, since auditing needs a guarantee.
\textbf{(C3) Scale.} Real assembly programs have $10^4$--$10^6$ gates and more than 10 qubits, beyond dense-matrix equivalence checking, while the general-purpose equivalence checkers that scale can run for hours or return inconsistent answers on large, differently-shaped circuits.
\textbf{(C4) Open-endedness.} The space of algorithms is unbounded. A recovery scheme built for a fixed list of algorithms misses every new family, and a fully generic one (rerolling loops from gate statistics) recovers little semantics.

\paragraph{Our approach.}
We present \sys, a decompiler for quantum assembly that checks each of its outputs for equivalence with the input.
\sys is organized around a single design invariant:
\begin{quote}\emph{Every transformation is either an exact rewrite checked where
it is applied, or a hypothesis checked against the input before use; anything
that fails its check is left in gate-level form.}\end{quote} \sys works in three
stages. It first \emph{lifts} the assembly back to the program's own qubits: it
removes the routing \texttt{SWAP}s, including those merged into other gates, by
relabeling wires (C1), and rewrites the remaining gates into a canonical form,
for example two-qubit blocks into the named gates they compute. Each of these
rewrites is checked where it is applied, so the lifted circuit is equivalent to
the input by construction (C2, C3). \sys then \emph{recovers} structure in two
ways. Top-down, \emph{algorithm sketches} are templates whose parameters are
computed from quantities that compilation leaves unchanged, instead of searched
for. Bottom-up, library recognition and loop rerolling alleviate the need of
algorithm-specific knowledge, so they also apply to algorithms that no sketch
anticipates (C4). Finally, \sys \emph{checks} the whole program against the
input.

\paragraph{Results.}
We built \textsc{QDecBench}, a benchmark of 816 compiled quantum assembly programs with ground-truth structure.
\sys's output is equivalent to the input for every program, and 94\% of the programs are fully decompiled, meaning that checked structure explains at least 80\% of their gates.
This is 1.7$\times$ as many as the best LLM-based decompiler, and \sys makes no model calls.
On nine algorithm families for which \sys has no sketch, recovered structure still explains 60\% of the gates.
The recovered programs also help downstream: an LLM identifies the algorithm more often from them than from the assembly (95\% against 68\% at the highest compilation level), and recompiling them for another device avoids the routing overhead that the assembly carries over.

\paragraph{Contributions.}
\begin{itemize}
  \item We formulate \emph{quantum assembly decompilation}, recovering a
  structured program equivalent to a compiled assembly program up to global
  phase and a reported output permutation (\S\ref{sec:background}).
  \item We design \sys around one principle: every step is checked. Exact
  rewrites undo the transformations of the compiler, and algorithm templates
  whose parameters are computed from the circuit recover the program's
  structure; a bottom-up path also recovers loops and library calls for
  algorithms without a template. These techniques make this robust to routing,
  native gates, and optimization, and let the check scale to millions of gates
  (\S\ref{sec:overview} and \S\ref{sec:design}).
  \item We implement \sys, build the \textsc{QDecBench} benchmark, and evaluate
  \sys against trivial, template-matching, and LLM-based baselines, with
  ablations and on real-world QASMBench files (\S\ref{sec:eval}).

\end{itemize}

\section{Background and Motivation}
\label{sec:background}

\subsection{Quantum Circuits and Assembly}
\label{sec:primer}
This subsection explains what a circuit means and what the programs we decompile look like. \citet{nielsen} give a full treatment, and Appendix~\ref{app:gates} lists the gates used in the paper with their circuit symbols.

\paragraph{Circuits and their meaning.}
The state of $n$ qubits is a vector of $2^n$ complex amplitudes, one per bit
string $x$ (written $|x\rangle$), and measuring the qubits returns $x$ with
probability equal to the squared magnitude of its amplitude. A gate is a unitary
matrix on one or two qubits: \texttt{cx} (CNOT) maps bit strings as \texttt{b
\textasciicircum= a}, \texttt{h} is the Hadamard gate, and rotations such as
\texttt{rz}$(\theta)$ take a real angle $\theta$. A circuit is a straight-line
program of gates, and its meaning $\sem{C}$ is the product of the gate matrices,
a $2^n\times 2^n$ unitary. Gates on disjoint qubits commute, so a circuit is a
DAG of gates, the \emph{commutation DAG}, similar to the dependence graph of a
basic block. Multiplying all amplitudes by one unit complex number $e^{i\theta}$
(a \emph{global phase}) changes no measurement outcome, so circuits are compared
up to a global phase and a reported permutation of the qubits
(\S\ref{sec:problem}). Quantum assembly spells a circuit out as text, one gate
application per line (\texttt{cx q[0],q[1];}); we use
OpenQASM~2~\cite{openqasm2}.

\paragraph{Source programs.}
Quantum programs are assembled from a small set of recurring building blocks.
The quantum Fourier transform (QFT) is the quantum analogue of the discrete
Fourier transform; quantum phase estimation (QPE) estimates an eigenphase of an
operator $U$ from the controlled powers $U, U^2, U^4, \ldots$ followed by an
inverse QFT; and Grover search finds a marked bit string by repeating an oracle
and a diffusion step. Many other algorithms are layers of rotations. Every
rotation gate is a \emph{Pauli rotation} $R_P(\alpha)=e^{-i\alpha P/2}$ about a
\emph{Pauli string} $P$, a tensor product of the Pauli matrices $X$, $Y$, $Z$
and the identity: \texttt{rz} is $R_Z$, \texttt{rx} is $R_X$, and \texttt{rzz}
is $R_{ZZ}$. QAOA alternates cost layers of $\texttt{rzz}(\gamma)$ gates over
the edges of a problem graph with mixer layers of $\texttt{rx}(\beta)$ gates,
and Trotterized Hamiltonian simulation and variational ans\"atze repeat similar
layers. A decompiler has to recover these building blocks together with their
parameters: the angles $\gamma$ and $\beta$, the eigenphase of a QPE, the marked
string of a Grover search.

\subsection{Quantum Circuit Compilation}
\label{sec:compile}
A quantum compiler maps a circuit onto what one device can execute, with passes
that mirror a classical back end; each pass preserves the meaning of the circuit
up to a global phase and, after routing, a qubit permutation
(Figure~\ref{fig:motivation}b). \emph{Basis translation} rewrites every gate
into the few native gates of the device, such as \texttt{rz}, \texttt{sx}, and
one two-qubit gate (\texttt{cx}, \texttt{ecr}, or \texttt{cz}). \emph{Layout and
routing} place the program's qubits on the device, where a two-qubit gate can
act only on qubits that are neighbours in the \emph{coupling map}; the router
inserts \texttt{SWAP} gates, each costing three \texttt{cx}, to bring qubits
together (in Figure~\ref{fig:motivation}, routing adds 27 of the 111 gates).
\emph{Optimization} then merges and cancels gates across the boundaries of the
program's building blocks, and a \texttt{SWAP} can merge into a neighbouring
two-qubit gate so that it no longer appears as a gate of its own. The compiler
emits the result as quantum assembly, together with the final
logical-to-physical permutation.

\paragraph{What compilation preserves.}
Since every pass replaces gates by others with the same meaning, a decompiler
must read structure from quantities that these rewrites leave unchanged. We use
three. (1)~\emph{Rotation angles.} Clifford gates (generated by \texttt{h},
\texttt{s}, and \texttt{cx}) change only the axis of a rotation they move past,
not its angle, so the angles of a program survive the Clifford gates that
compilation interleaves with them (\S\ref{sec:pauli}). (2)~\emph{The interaction
of a two-qubit block.} Every two-qubit unitary needs at most three
CNOTs~\cite{vatan}, and three numbers, its \emph{Weyl-chamber
coordinates}~\cite{weyl}, describe its interaction regardless of the
single-qubit gates around it; they reveal \texttt{SWAP}s hidden inside other
blocks (\S\ref{sec:unroute}). (3)~\emph{The truth table of classical regions.}
Gates such as \texttt{x}, \texttt{cx}, and the Toffoli gate permute bit strings,
and phase gates multiply each bit string by a phase. A region built from them,
which we call \emph{monomial} (its matrix has one nonzero entry per column), is
fully described by this table, whichever gates implement it (\S\ref{sec:canon}).

\subsection{Problem: Decompiling Quantum Assembly}
\label{sec:problem}
Given a compiled quantum assembly program $C$ with no metadata, the decompilation problem is to produce a program $P$ in a high-level host language (here, Python/Qiskit with loops, functions, library operations, and symbolic angles) such that
(i) \emph{equivalence}: $P\eqv C$, that is, $\sem{P} = e^{i\theta}\,\Pi\,\sem{C}$ for some global phase $\theta$, where $\Pi$ is the wire permutation that $P$ reports (the final logical-to-physical map, as compilers report it), and each wire is measured into the same classical bit in $P$ as its image under $\Pi$ is in $C$. Because $\Pi$ is reported, the equation fixes the action on every wire, measured or not; and
(ii) \emph{structure}: $P$ makes the algorithmic structure of $C$ explicit: $P$ is much shorter than $C$ and its statements name what they compute.
Only (i) is a hard requirement. A program that lists $C$ gate by gate satisfies it trivially, so (ii) is a best-effort objective, measured by the metrics of \S\ref{sec:eval}.

\paragraph{Comparison with classical decompilation.}
\label{sec:cmp}
The goal is that of C/C++ decompilers, but three differences shape our design.
First, there is an exact semantic oracle. No cheap check confirms that a decompiled C program matches its binary~\cite{cifuentes,phoenix}, whereas two circuits can be checked for equivalence exactly, by comparing matrices for up to 10 qubits and with checkers based on decision diagrams or the ZX-calculus beyond~\cite{qcec}. Recovery can therefore propose hypotheses and keep only those that pass the check.
Second, the main artifacts of compilation, such as routing \texttt{SWAP}s, basis-translation identities, and merged gates, change the syntax of the program but not its meaning, so exact local rewrites can remove them.
Third, quantum programs are dominated by real-valued parameters such as rotation angles, and the compiler folds them together with \emph{gauge freedoms}, different assignments that yield the same circuit; for example, an \texttt{rz} on the control qubit of a \texttt{cx} can sit on either side of it. Recovery must therefore solve for parameters modulo gauge.

\paragraph{LLM-based decompilation}
\label{sec:llm}
A large language model can be asked to turn the QASM text into a Qiskit
program~\cite{llm4decompile,decllm}. On compiled assembly, this gives no
guarantee: even with a repair loop that feeds back counterexamples, the model
must transcribe hundreds to thousands of gates and often produces an
inequivalent or merely transcribed program (\S\ref{sec:eval}). Much of the
structure is also numerical, such as the marked state of a routed Grover search
or the angles of a QAOA circuit whose mixers were merged into neighbouring
rotations, and cannot be read from the gate list. \sys therefore recovers
structure with exact rewrites, solvers over circuit invariants, and equivalence
checks, without a language model.

\section{Overview and Design Considerations}
\label{sec:overview}

\subsection{Design Considerations}
Following the differences identified in \S\ref{sec:cmp}, we require a quantum
decompiler to satisfy: \textbf{R1 (Soundness).} Never report a program as
equivalent unless it is; failing to recover structure degrades the output
without corrupting it. \textbf{R2 (Generalizability).} Do not assume the
compiler, its optimization level, or the coupling map; the same algorithm
compiled differently must yield the same program. \textbf{R3 (Scalability).}
Handle $10^5$-gate, $>\!10$-qubit inputs, and degrade to \texttt{unverified}
within a time budget. \textbf{R4 (Automation and best-effort structure).} Run
without user hints, and recover as much loop, function, and library structure as
can be justified.

\subsection{Workflow}

\begin{figure}[t]
\centering
\includegraphics[width=\textwidth]{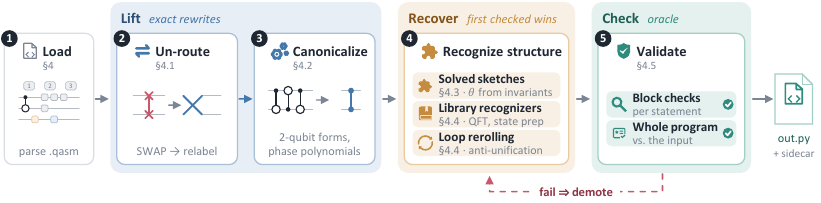}
\caption{\sys's pipeline.}
\label{fig:pipeline}
\end{figure}

Figure~\ref{fig:pipeline} shows \sys's pipeline. The input is a compiled quantum
assembly program (OpenQASM); the output is a Python/Qiskit module exposing
\texttt{build\_circuit()}, plus a \emph{sidecar} that records each recovered
structure (type, qubits, iteration counts, children, the input gates it
explains), the structural coverage, and the final logical-to-physical mapping.
\sys includes the following stages: \textbf{Load} (\S\ref{sec:design}) parses the
assembly into a circuit, expands custom gates, and tags every gate with an
identifier so that provenance can be tracked; then, \textbf{Lift}
(\S\ref{sec:unroute}--\ref{sec:canon}) removes compilation artifacts with exact
rewrites: \emph{un-routing}, two-qubit canonicalization, single-qubit merging,
monomial-region summarization, H-con\-ju\-ga\-tion lifting, and \texttt{SWAP}
virtualization; \textbf{Recover} (\S\ref{sec:sketch}--\ref{sec:reroll}) builds a
structured \emph{program IR} from the lifted circuit by solved sketches, or by
library recognition and loop rerolling; and finally, \textbf{Generate and check}
(\S\ref{sec:verify}) renders the IR as code, checks each structured statement
against the operations it replaces, and checks the whole program against the
input.

\subsection{The Design Invariant}
\label{sec:invariant}
Let $\eqv$ denote equivalence up to global phase and a tracked wire permutation (\S\ref{sec:problem}).
\sys is built from two kinds of steps only.
An \emph{exact rewrite} replaces a region of the circuit by another and is applied only if a local check shows that the two are equivalent (dense unitaries for small blocks, value tables for monomial regions).
A \emph{hypothesis} proposes a structured program for the lifted circuit $L$ and is used only if a check shows that it is equivalent to $L$.

\paragraph{Soundness.}
For every input $C$, \sys either emits a program $P\eqv C$ and reports it as \texttt{equivalent}, or emits the gate-level program of the lifted circuit and reports it as \texttt{unverified}; up to the trusted base below, it never reports \texttt{equivalent} for a program that is not equivalent to $C$.
The argument follows the pipeline. A rewrite that fails its check is not applied, and a hypothesis that fails its check leaves its gates in place, so every intermediate program is equivalent to the input.
The final program is checked against the input once more; if this check fails, \sys emits and checks the gate-level program instead, and if a check runs out of time, it reports \texttt{unverified}.

\paragraph{Trusted base.}
Up to 10 qubits, the final check compares $P$ with the input itself, so it relies only on the library that builds and multiplies the unitaries (Qiskit).
Beyond 10 qubits, it compares $P$ with the lifted circuit (\S\ref{sec:verify}), and the trusted base also includes the bookkeeping that composes the local checks of lifting, such as the relabeling map of un-routing that fixes the final order of the wires; the H-conjugation identity, which is not checked for regions wider than 10 qubits; and MQT QCEC, whose verdict counts only when it reports equivalence.

The invariant has two consequences for the rest of the paper.
First, the equivalence of the output does not depend on whether recovery succeeds: operations that no checked structure explains are emitted as gate-level code, which is already easier to read than the input because routing is removed and gates are merged.
Second, recovery techniques can propose candidates that are fitted numerically or enumerated, such as rotation angles or iteration counts, because the check rejects every incorrect candidate.

\section{Design}
\label{sec:design}

This section follows the pipeline of Figure~\ref{fig:pipeline}. \sys parses the input into a list of instructions, expands custom gates, and gives each instruction an identifier; every later structure records the identifiers of the input gates it explains (its \emph{provenance}), from which the sidecar and SC are computed. Lifting removes routing (\S\ref{sec:unroute}) and rewrites the remaining gates into canonical forms (\S\ref{sec:canon}); recovery builds structure top-down with solved sketches (\S\ref{sec:sketch}) and bottom-up with library recognition and loop rerolling (\S\ref{sec:reroll}); \S\ref{sec:verify} checks the result, and \S\ref{sec:impl} describes the implementation.

\subsection{Un-Routing}
\label{sec:unroute}
\begin{figure}[t]
\centering
\includegraphics[width=\textwidth]{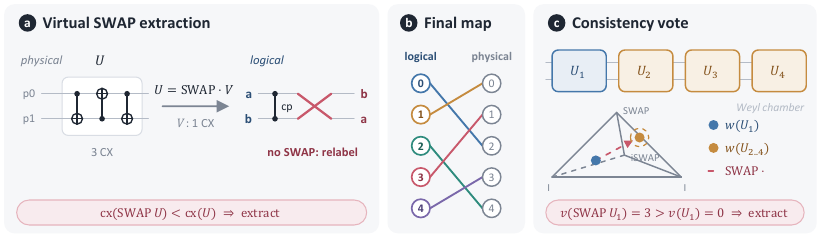}
\caption{Un-routing.}
\label{fig:unroute}
\end{figure}
Routing is the dominant artifact of compilation and the main obstacle to recovery: every sketch reads parameters from the \emph{logical} circuit, so a permuted, \texttt{SWAP}-laden circuit defeats them all (Figure~\ref{fig:ablation}).
Un-routing recovers the logical circuit and the routing permutation.

\paragraph{Two-qubit blocks.}
Gates are grouped into maximal two-qubit blocks; grouping depends only on qubit connectivity, not labels, so blocks are collected first and decisions are replayed over the event stream.
For a block with unitary $U\in\U(4)$, let $\mathrm{cx}(U)\in\{0,1,2,3\}$ be its minimal CNOT count, computed from its Weyl-chamber coordinates $w(U)$~\cite{weyl,vatan}, the three numbers that describe $U$ up to single-qubit gates on either side (\S\ref{sec:compile}).

\paragraph{Virtual \texttt{SWAP} extraction.}
If $\mathrm{cx}(\textsf{SWAP}\cdot U) < \mathrm{cx}(U)$, the block is rewritten as $(\textsf{SWAP}\cdot U);\textsf{SWAP}$ and the trailing \texttt{SWAP} is not emitted; the logical labels of the two wires are exchanged instead (a virtual relabeling, Figure~\ref{fig:unroute}a).
The running map is threaded through the rest of the circuit, and the final logical-to-physical permutation (Figure~\ref{fig:unroute}b) and the measurement map are updated accordingly and later checked (\S\ref{sec:verify}).

\paragraph{Consistency vote for cost-ambiguous blocks.}
A \texttt{SWAP} merged into a generic two-qubit gate (one that needs three CNOTs) leaves the count at three, so the cost criterion cannot see it.
We exploit that programs \emph{repeat} their interactions (Trotter steps, ansatz layers, QAOA layers).
Weyl coordinates ignore the single-qubit gates the compiler merges around a block, so the coordinates $w(U)$ recur across repetitions of the same logical interaction, whereas $w(\textsf{SWAP}\cdot U)$ recurs when the routing swapped consistently.
The block's \texttt{SWAP} is extracted when the coordinates of $\textsf{SWAP}\cdot U$ occur elsewhere in the circuit more often than those of $U$ (Figure~\ref{fig:unroute}c).
\begin{equation}
\mathrm{extract}(U)\iff\mathrm{cx}(\textsf{SWAP}\,U)<\mathrm{cx}(U)\;\vee\;\big(\mathrm{cx}(\textsf{SWAP}\,U)=\mathrm{cx}(U)\wedge v(\textsf{SWAP}\,U)>v(U)\big),
\end{equation}
where $v(V)$ counts the other blocks of the circuit whose Weyl coordinates equal $w(V)$.
Each extraction is an exact rewrite: the $4\times 4$ unitary of the block before and after is compared (\S\ref{sec:verify}).

\paragraph{Canonical two-qubit forms.}
After extraction, each block is re-expressed as $(\ell_1\otimes\ell_2)\cdot\mathit{core}\cdot(\ell_3\otimes\ell_4)$, a two-qubit $\mathit{core}$ with single-qubit gates $\ell_i$ on each side. The core is chosen from named families in order of preference: diagonal (\texttt{cp}, \texttt{rzz}), controlled (\texttt{cx}, \texttt{cz}, \texttt{cry}, \texttt{cu}), canonical interaction ($\texttt{rxx}\cdot\texttt{ryy}\cdot\texttt{rzz}$), Pauli rotations, and finally the generic KAK form that every two-qubit unitary admits~\cite{vatan}.
Single-qubit factors are factored as $P\cdot\mathit{core}\cdot P$ into named gates (\texttt{h}, \texttt{s}, \texttt{t}, \texttt{sx}, \texttt{p}, \texttt{rx}, \texttt{ry}, \texttt{rz}).
This is what turns three \texttt{cx} and eight \texttt{rz}/\texttt{sx} back into one \texttt{cp(\ldots)} (Figure~\ref{fig:canon}a).

\paragraph{Interaction frames.}
Which local gates surround a core depends on the device's native two-qubit gate. A compiler that targets \texttt{cz} or \texttt{ecr} wraps every entangling gate in the single-qubit gates that turn it into the program's interaction and merges them with the program's own rotations, so the same interaction reaches the lifted circuit in different local frames: a $ZZ$ rotation compiled to \texttt{cz} gates, for example, appears as a controlled-$X$ rotation whose angle leaks into a neighbouring \texttt{rx}.
Every core whose interaction is a single Pauli-product rotation (the CNOT class included) can be written as $L_1 D L_0$ with a diagonal $D$ and local $L_0, L_1$, which are Clifford gates whenever the core conserves one local Pauli axis on each qubit, as native gates and controlled Pauli rotations do.
Two choices leave the core unchanged: its \emph{frame} (the named core, \texttt{cz} or \texttt{cx} in either orientation for the CNOT class and \texttt{cp} or \texttt{rzz} for a rotation, together with the Pauli part of $L_0$ and $L_1$) and its \emph{gauge} (a rotation about a conserved axis commutes with the core and may move from one side of it to the other).
\sys fixes both so that the merged single-qubit runs are as simple as possible, counting each run that leaves the monomial fragment (\S\ref{sec:canon}) and each non-Clifford rotation.
The choices of consecutive cores on a qubit interact through the run between them, so they are optimized one qubit line at a time by dynamic programming (Viterbi), and the result is kept only if it lowers the total.
Each alternative form is checked against the core's $4\times 4$ unitary before it is offered and again before it is emitted.
Template fitting (\S\ref{sec:fit}) compares entangling skeletons modulo the same local Clifford frames.

\paragraph{Initial layout.}
Un-routing undoes the permutations that routing applies \emph{during} the circuit, but not the compiler's \emph{initial layout}, a renaming of the program's qubits that no gate records: the program's qubit $i$ starts on some wire $\pi(i)$.
Structure that depends on the order of a register (the carry chain of an adder, the entangling order of an ansatz) is invisible under the identity naming.
We treat $\pi$ as one more parameter to solve.
Measurements survive compilation, so the classical bit that each wire is measured into orders the register (unmeasured wires follow); when recovery under the identity naming yields no sketch, recovery (\S\ref{sec:sketch}--\ref{sec:reroll}) runs again on the circuit relabeled by $\pi$.
A result on the relabeled circuit is emitted as a program on the program's qubits, composed onto the wires $\texttt{LAYOUT}=\pi$; since the composed program acts on the input's wires, the checker is unchanged.

\subsection{Canonicalization}
\label{sec:canon}
\begin{figure}[t]
\begin{minipage}[t]{0.485\textwidth}
\centering
\includegraphics[width=\linewidth]{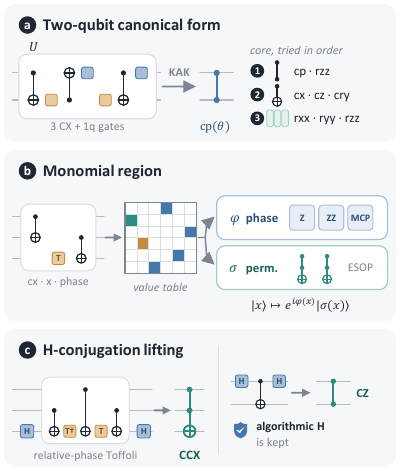}
\caption{Canonicalization rewrites.}
\label{fig:canon}
\end{minipage}\hfill
\begin{minipage}[t]{0.485\textwidth}
\centering
\includegraphics[width=\linewidth]{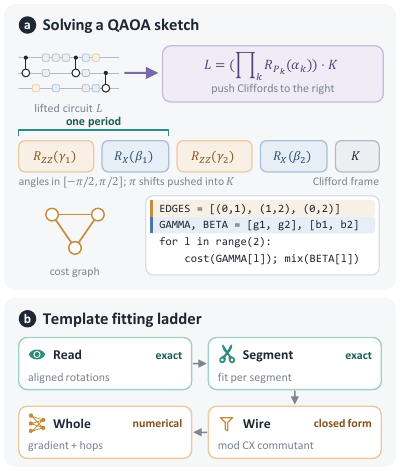}
\caption{Solved sketches.}
\label{fig:sketch}
\end{minipage}
\end{figure}
\paragraph{Single-qubit merging.}
Runs of single-qubit gates are merged. We keep two views of the circuit: a \emph{canonical} view for structure matching and a \emph{light} view that preserves per-qubit rotation structure for template fitting (\S\ref{sec:sketch}).

\paragraph{Monomial regions.}
A region consisting of \texttt{cx}, \texttt{x}, and phase gates is monomial (\S\ref{sec:compile}): it maps $|x\rangle\mapsto e^{i\varphi(x)}|\sigma(x)\rangle$, a permutation $\sigma$ of bit strings with a phase function $\varphi$.
We select regions that are \emph{convex} in the commutation DAG (no dependence path leaves a region and re-enters it), so a region can be replaced without reordering other gates.
Each region is summarized by its value table, the pair $(\sigma(x),\varphi(x))$ for every input $x$, which is feasible up to 16 qubits (Figure~\ref{fig:canon}b).
The phase function $\varphi$ is rewritten as a polynomial over the input bits in its sparsest basis.
Products of bits (monomials) become multi-controlled phases, and indicator functions of one bit pattern become multi-controlled phases conjugated by \texttt{x}.
XORs of bits (parities) become $Z/ZZ$ rotations, followed by a clean-up over GF(2) (bits under XOR) in the style of phase-polynomial optimization~\cite{amy,nam}.
The permutation $\sigma$ is resynthesized as an ESOP, an XOR of AND terms with one multi-controlled X gate per term, or, if it is affine, as a network of \texttt{cx} gates.

\paragraph{H-conjugation lifting.}
Placing Hadamards on both sides of a region $R$ on a qubit $t$ exchanges the roles of $X$ and $Z$ on $t$ ($HZH=X$). Where $R$ acts on $t$ only through Paulis, $H_tRH_t$ is lifted to a single operation. This recovers Toffoli and multi-controlled X gates from the relative-phase decompositions compilers emit, and \texttt{cz} from \texttt{h;cx;h} (Figure~\ref{fig:canon}c).
Solitary Hadamards that belong to the algorithm (e.g., a state-preparation layer) are protected.

\paragraph{Wire permutations and \texttt{SWAP} virtualization.}
A region of \texttt{cx} and \texttt{x} gates computes an affine map of the input bits, $x\mapsto Mx\oplus c$ with an invertible bit matrix $M$. We factor $M=P\cdot L$ where $L$ has a unit diagonal (a perfect matching of ones, which exists because $M$ is invertible); $L$ becomes \texttt{cx} gates and $P$ explicit \texttt{SWAP}s.
\texttt{SWAP}s, $\texttt{cx}(a,b)\texttt{cx}(b,a)\texttt{cx}(a,b)$ triples, and partially cancelled pairs $\texttt{cx}(a,b)\texttt{cx}(b,a)=\texttt{cx}(b,a)\texttt{swap}(a,b)$ are then removed by relabeling.
Because a program's \emph{own} CX pair can look like a routing residue, recovery is also run on the un-virtualized circuit whenever the virtualized one yields no sketch, and the better program is kept.

\paragraph{Semantic collapse.}
Windows whose unitary is monomial after peeling a constant X layer are replaced by their synthesized classical function.

\subsection{Solved Sketches}
\label{sec:sketch}
After lifting, the circuit $L$ is close to its logical form. Recovery turns $L$ into a \emph{program IR} of gates, loops, and code blocks, each with provenance.
The top-down strategy uses \emph{sketches}: a sketch is a parameterized algorithm template $S(\theta)$ together with a \emph{solver} $\sigma$ that computes $\theta=\sigma(L)$ from invariants of $L$, rather than searching for $\theta$ by enumeration or gradient-free optimization as in classical sketch-based synthesis~\cite{sketch}.
The hypothesis $S(\sigma(L))$ is then checked against $L$ (a random-state probe as a prefilter, then the exact check) and the first hypothesis that passes the check wins.
Sketches are tried in a fixed order, cheap and specific first (Appendix~\ref{app:sketches} lists them). Preconditions on qubit count and gate kinds keep failed attempts inexpensive.

\subsubsection{Pauli-rotation normal form}
\label{sec:pauli}
QAOA and Trotterized Hamiltonian simulation are, at the source level, sequences of Pauli rotations (\S\ref{sec:primer}); compilation interleaves them with Clifford gates that change each rotation's apparent axis. They are recovered by rewriting $L$ into
\begin{equation}
L \;=\; \Big(\prod_{k} R_{P_k}(\alpha_k)\Big)\cdot K,
\end{equation}
a sequence of Pauli rotations followed by a single Clifford circuit $K$ (the \emph{Clifford frame}), by pushing Cliffords to the right with the rule $K\,R_P(\alpha)=R_{KPK^\dagger}(\alpha)\,K$.
Rotations with Clifford angles join the frame; angles are reduced to $[-\pi/2,\pi/2]$ by pushing $R_P(\pi)=-iP$ into the frame; and $\pi/2$ (S-type) gauges that the compiler merged into rotations are fixed by a consistency vote over the recurring Pauli.
The period of the normalized rotation sequence gives the number of layers; the angles give $\gamma$ and $\beta$ (Figure~\ref{fig:sketch}a).
For QAOA, angles outside the window are recovered from the frame bookkeeping: a cost layer's Pauli anticommutes exactly with the mixers of odd-degree vertices, so the sign pattern of later mixer layers reveals each layer's $\pi$ shift.

\subsubsection{Template fitting modulo gauge}
\label{sec:fit}
For variational ans\"atze and QNNs the template (library \texttt{real\_amplitudes}, \texttt{efficient\_su2}, feature maps) fixes the CX skeleton but not the rotation angles, which the compiler has merged into neighbours.
We first compare the CX skeleton per qubit modulo commutation of same-role CX runs. Parameters are then read from aligned rotations; if that fails, fitted per segment; then per wire modulo the CX commutant, where gauges ($R_z$ on controls and $R_x$ on targets commute through CX) are eliminated in closed form; and finally fitted on the whole circuit with analytic gradients (one forward and one backward statevector sweep over random probes) with coordinate-wise Clifford-step hopping ($\pm\pi/2,\pi$) out of shallow minima.
This \emph{fitting ladder} (Figure~\ref{fig:sketch}b) climbs from exact reads to numerical fits, so the cheap, exact rung is used whenever it applies. The result is checked exactly, so a fit that converges to the wrong basin is rejected.

\subsection{Library Recognition and Loop Rerolling}
\label{sec:reroll}
\begin{figure}[t]
\begin{minipage}[t]{0.485\textwidth}
\centering
\includegraphics[width=\linewidth]{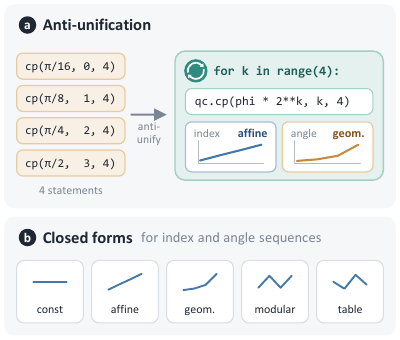}
\caption{Loop rerolling.}
\label{fig:reroll}
\end{minipage}\hfill
\begin{minipage}[t]{0.485\textwidth}
\centering
\includegraphics[width=\linewidth]{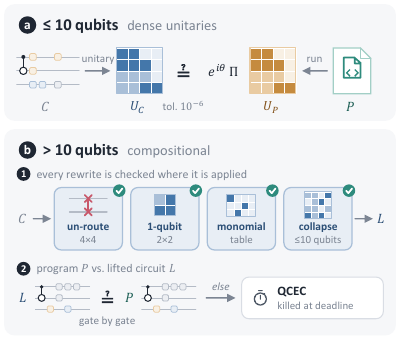}
\caption{Equivalence checking.}
\label{fig:verify}
\end{minipage}
\end{figure}
The bottom-up strategy recovers structure from the lifted circuit itself, without an algorithm template.
\emph{Library recognizers} (QFT and inverse QFT chains, basis and uniform state preparation) are contracted only when they are convex in the commutation DAG and pass their check.
\emph{Loop rerolling} finds runs of repeated statements and anti-unifies~\cite{plotkin} their templates (Figure~\ref{fig:reroll}): index and angle sequences are fitted by closed forms (constant, affine, geometric with ratio 2 or $\tfrac12$, modular, table), which yields nested loops, zipped loops over edge lists (\texttt{for a, b in EDGES}), broadcast layers (\texttt{qc.h(range(n))}), and loop extension by closed-form prediction.
A loop is kept only if its block check passes.
Because this path needs no algorithm-specific knowledge, it also decompiles algorithms that no sketch anticipates to a level of abstraction well above the gate list: loops, broadcast layers, library calls such as the QFTs inside HHL, and multi-controlled gates; on nine unseen families, checked structure explains 60\% of the gates (\S\ref{sec:rq2}).
Two statement orders (canonical ASAP and source order) are tried, and the program with the smaller rendered size wins.

\subsection{Equivalence Checking}
\label{sec:verify}
\paragraph{Block checks.}
Every structured statement is checked against the operations it replaces: dense operators for $\le 10$ qubits, random-state stimuli beyond. A statement that fails is demoted to its gates.

\paragraph{Whole-program check.}
The program is executed in a fresh interpreter (so that generated code cannot influence the checker) and checked against the input.
For $\le 10$ qubits, we compare the measurement map and the unitaries (dense operators, absolute tolerance $10^{-6}$ after global-phase alignment) with the final permutation applied (Figure~\ref{fig:verify}a).
Beyond 10 qubits, dense comparison is infeasible and general equivalence checkers can time out or, on structurally different circuits, disagree. \sys therefore checks it \emph{compositionally} (Figure~\ref{fig:verify}b):
\begin{enumerate}
  \item every lifting rewrite is validated where it is applied (translation validation~\cite{pnueli}): $4\times4$ block unitaries in un-routing, $2\times2$ single-qubit factorizations, value tables for rewritten monomial regions, dense unitaries for collapsed windows and H-conjugated regions up to 10 qubits; a rewrite that fails is not applied. Hence the lifted circuit $L$ is equivalent to the input up to the tracked permutation.
  \item it remains to check the program $P$ against $L$. Since $P$ and $L$ are \emph{structurally aligned} (the program is built from $L$), gate-by-gate comparison decides the check exactly in linear time; otherwise we fall back to MQT QCEC~\cite{qcec}, an equivalence checker based on decision diagrams.
\end{enumerate}
QCEC runs in a child process that is killed at its deadline (its own timeout is cooperative and can overrun by hours), after random-state stimuli have turned any disagreement into a counterexample. A QCEC verdict of ``not equivalent'' triggers the fallback to the flat canonical program and is never reported as final. A check that runs out of time reports \texttt{unverified}.

\subsection{Implementation and Usage}
\label{sec:impl}

\sys is implemented in about 7.0K lines of Python 3.12 on top of Qiskit~\cite{qiskit}, NumPy, and MQT QCEC~\cite{qcec}; the lifting engine accounts for 6.6K of them. The LLM client used by the baselines caches every request, so all LLM runs can be reproduced offline.
Each stage receives a share of the time budget, at most a quarter for lifting and half for recovery. A stage that reaches its share stops refining and passes on the equivalent circuit it has, so a very large input yields a program with less structure; otherwise the engine is deterministic.
Unlike the decompiler of Liu et al.~\cite{btd}, whose recovery errors may need manual repair, \sys needs no repair step (\S\ref{sec:invariant}). A user can inspect each recovered structure in the sidecar, together with its provenance and block check, and re-run the check independently.

\section{Evaluation}
\label{sec:eval}

We evaluate \sys along five research questions:
\begin{itemize}
  \item \textbf{RQ1 (Soundness and effectiveness).} Is every output equivalent to its input, and how much structure does \sys recover compared with the baselines and with LLM-based decompilers?
  \item \textbf{RQ2 (Robustness and generalization).} Does recovery survive changes of the compilation pipeline, and does it extend to algorithms outside the benchmark?
  \item \textbf{RQ3 (Contribution of techniques).} Which techniques matter?
  \item \textbf{RQ4 (Scalability and real-world assembly).} Does \sys scale to large inputs and to assembly it was not designed on?
  \item \textbf{RQ5 (Downstream applications).} Does the recovered program help a reader understand the assembly, and does it help retarget the assembly to another device?
\end{itemize}

\subsection{Setup}
\label{sec:setup}
\paragraph{Benchmark.}
No benchmark of compiled assembly with ground-truth structure exists, so we
build \textsc{QDecBench} (816 assembly programs). Built from MQT
Bench~\cite{mqtbench} or Qiskit library constructions with random parameters, it
has 22 algorithm families in three tiers, from textbook kernels (\textbf{T1})
through algorithms (\textbf{T2}) to composite circuits (\textbf{T3}), and random
circuits as a \emph{negative control}; Appendix~\ref{app:bench} describes them.
Each program is compiled with Qiskit at four levels that move progressively
further from the source: \textbf{C0} and \textbf{C1} translate it to a hardware
gate set (C1 also optimizes), and \textbf{C2} and \textbf{C3} additionally place
and route it on a device coupling map, with moderate and aggressive
optimization, for one of three native two-qubit gates: \texttt{cx}, \texttt{ecr}
(IBM Eagle), or \texttt{cz} (IBM Heron); Appendix~\ref{app:levels} gives the detailed settings.

\paragraph{Metrics.}
\textbf{ER} (equivalence rate): the emitted program executes and is equivalent to the input. \textbf{SC} (structural coverage): the fraction of input gates attributed, through provenance, to checked structures (routing \texttt{SWAP}s and cancelled gates count as explained). \textbf{FD} (full decompilation): ER and SC $\ge 0.8$. \textbf{Sub-F1}: F1 of recovered structure types against the ground truth. \textbf{LR} (loop recovery): fraction of ground-truth loops recovered with the exact iteration count. \textbf{CR} (compression ratio): input gates per output statement.
LLM baselines report no provenance that could be checked, so for them we use \textbf{SF} instead of SC: the fraction of the output's gates that the program emits from loops, reused functions, gate layers, or library calls, measured by tracing its execution (Appendix~\ref{app:sf}); FD(SF) is ER and SF $\ge 0.8$.

\paragraph{Baselines.}
\leavevmode\ding{192}~\emph{Trivial-GbG}: the input transcribed gate by gate.
\ding{193}~\emph{Trivial-TM}: we match a library of templates for common
subroutines directly against the assembly, without lifting
(Appendix~\ref{app:tm}). \ding{194}~\emph{DirectLLM}: two reasoning models,
DS-Flash (\texttt{deepseek-flash}) and GPT-6.1-Sol (\texttt{gpt-6.1-sol}),
both with high reasoning effort and a 64K-token output limit, receive the raw
QASM and must return a Qiskit program and its structure list.
\ding{195}~\emph{DecLLM}~\cite{decllm}: DirectLLM followed by DecLLM's repair
loop, adapted to quantum assembly: up to three further rounds that feed back
exceptions, dimension or measurement mismatches, or a counterexample basis
input. Each LLM configuration is run three times to account for non-determinism.
We also considered DeQompile~\cite{deqompile}, but it does not apply to a single
compiled program: it evolves a generator for one algorithm's circuits at several
sizes and scores it by textual similarity (see analysis in
Appendix~\ref{app:deqompile}).

\paragraph{Platform.}
All runs use an Intel Core Ultra~9 285K (24 cores) with 64\,GB RAM, Python~3.12, Qiskit~2.5.2, MQT QCEC~3.10.0, and MQT Bench~2.3.0. Engine runs are deterministic.

\subsection{RQ1: Soundness and Effectiveness}
\label{sec:rq1}

\paragraph{Soundness.}
Across all 816 inputs, no output of the 21 engine configurations, including every ablation, or of Trivial-TM is inequivalent to its input (ER $=1.000$), as the argument of \S\ref{sec:invariant} requires. The check is not vacuous. We mutate every output, for example by perturbing an angle, moving a gate, or exchanging two measurement targets, and classify each of the 12{,}366 mutants independently by statevector simulation: the checker rejects every mutant that changes the semantics and accepts the 5 that do not. Beyond 10 qubits, where outputs are checked compositionally (\S\ref{sec:verify}), we mutate the 12 QASMBench outputs checked this way (RQ4). No mutant that changes a gate, an angle, or a measurement passes; the only accepted mutants (13 of 36) exchange the final order of two unmeasured wires, which is part of the trusted base (\S\ref{sec:invariant}).

\paragraph{Structure recovery (Table~\ref{tab:main}).}
\sys fully decompiles 93.8\% of the inputs and recovers 98.0\% of the ground-truth loops with their exact iteration count. Compilation costs little: from C0 to C3, which adds routing, a transpiler-chosen initial layout, and aggressive optimization, FD falls by only 6\%. The hardest native gate is \texttt{ecr}, about 5 FD points below \texttt{cx} and \texttt{cz} at C2/C3. Template matching on the assembly (Trivial-TM) fully decompiles only 4.5\% of the inputs, mostly GHZ states and unrouted QFTs, because compilation fuses and interleaves the gates of a library operation, and merged angles and routing break the exact repetition that templates need.

\newsavebox\maintab
\sbox\maintab{\footnotesize\setlength{\tabcolsep}{2.6pt}\renewcommand{\arraystretch}{1.2}\begin{tabular}{@{}lrrrrrr@{}}
\toprule
 & ER & SC & FD & Sub-F1 & LR & CR\\
\midrule
\sys (all)   & 1.000 & .950 & .938 & .904 & .980 & 17.4\\
\quad C0 & 1.000 & .981 & .978 & .926 & 1.000 & 8.8\\
\quad C1 & 1.000 & .957 & .956 & .926 & 1.000 & 10.6\\
\quad C2 & 1.000 & .941 & .926 & .895 & .966 & 21.8\\
\quad C3 & 1.000 & .940 & .919 & .892 & .974 & 20.6\\
Trivial-GbG & 1.000 & 0 & 0 & .015 & 0 & 0.9\\
Trivial-TM & 1.000 & .055 & .045 & .062 & .057 & 0.9\\
\bottomrule
\end{tabular}}
\begin{figure}[t]
\begin{minipage}[t]{\wd\maintab}
\vspace{0pt}
\captionof{table}{Main results on \textsc{QDecBench}.}
\label{tab:main}
\usebox\maintab
\end{minipage}\hfill
\begin{minipage}[t]{0.55\textwidth}
\vspace{0pt}
\includegraphics[height=\dimexpr\ht\maintab+\dp\maintab\relax]{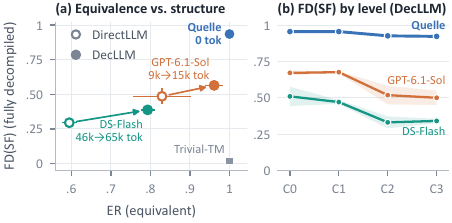}
\caption{Comparison with LLM-based decompilers.}
\label{fig:llm}
\end{minipage}
\end{figure}

\paragraph{Versus LLMs (Figure~\ref{fig:llm}).}
The LLM baselines run on all 816 inputs, three times each. On \sys's outputs, FD(SF) and FD agree ($0.935$ and $0.938$), so SF measures structure as SC does. The best LLM configuration, DecLLM with GPT-6.1-Sol, produces an equivalent program for 96.1\% of the inputs, but its FD(SF) of $0.564$ is 40\% below that of \sys, and it spends 15k tokens per input. Repair mainly raises equivalence (Figure~\ref{fig:llm}a): the models restore it largely by transcribing gates, so the share of structured code grows little. The LLMs also lose structure as compilation reshapes the circuit. From C0/C1 to C2/C3, FD(SF) of the best configuration falls by about a quarter, while \sys stays between $0.92$ and $0.96$ at every level (Figure~\ref{fig:llm}b).

\paragraph{Failure analysis.}
The 51 inputs that \sys does not fully decompile fall into three groups. Twenty-one are Deutsch--Jozsa programs with a constant oracle: the Hadamard layers around the oracle cancel, and the few gates left do not determine the algorithm. Twelve are the random circuits of the negative control, for which \sys claims no algorithm, as it does for 624 further random circuits (Appendix~\ref{app:bench}). The other 18 are partial recoveries at C2/C3, mostly compiled for \texttt{ecr} or \texttt{cz}, where no template or sketch matches; the output keeps the recognized parts, such as QFTs and Trotter terms, as loops and library calls.

\subsection{RQ2: Robustness and Generalization}
\label{sec:rq2}
\begin{figure}[t]
\centering
\includegraphics[width=\textwidth]{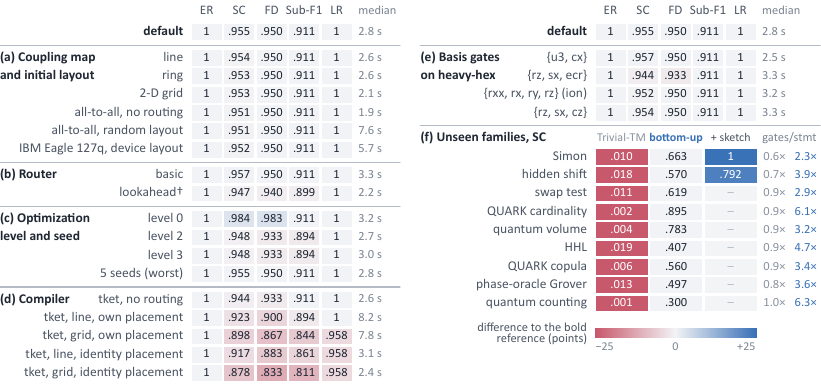}
\caption{Robustness across compilation conditions (a--e) and on algorithm families outside \textsc{QDecBench} (f); in (f), gates per statement of Trivial-TM (grey) and of the bottom-up path (blue).}
\label{fig:conditions}
\end{figure}

To test that recovery does not depend on one compilation pipeline (R2), we recompile 60 of the source programs under 31 further conditions and check each compiled program against its source (Figure~\ref{fig:conditions}). Three of these programs are never fully decompiled (the two constant-oracle Deutsch--Jozsa programs and the random circuit), so FD is at most $0.950$.

In Figure~\ref{fig:conditions}, the bold row at the top of each column is the default compilation (Qiskit, heavy-hex map, SABRE routing, level~1, seed~7, CX basis, identity initial layout), and a tile's colour shows its difference to that row; in (f), the reference is the bottom-up column.
\textbf{\textit{Coupling map and initial layout (a).}} No map or layout changes FD, Sub-F1, or LR, including the layout that the transpiler chooses on a 127-qubit IBM Eagle map, because \sys recovers the register order from the measurements (\S\ref{sec:unroute}).
\textbf{\textit{Router (b).}} The router has no effect either; the lookahead router compiles only 50 of the programs, so its FD is measured on those.
\textbf{\textit{Optimization level and seed (c).}} Levels 2 and 3 lose one program and level~0 gains two; the five transpiler seeds change the assembly of 34 programs but no metric.
\textbf{\textit{Compiler (d).}} tket is the largest change, with FD up to 12\% lower. Every loss comes from tket's \texttt{FullPeepholeOptimise}, which resynthesizes the two-qubit structure that the templates of ans\"atze, adders, and the entangled QFT rely on.
\textbf{\textit{Basis gates (e).}} The native gate has little effect: with \texttt{cz}, \texttt{ecr}, or the trapped-ion \textsf{rxx}, FD stays within 2 points of \texttt{cx}. Compilers for these gates express the same interaction in different local frames, and the interaction frames of \S\ref{sec:unroute} bring it back to one diagonal form (RQ3).

\paragraph{Algorithms outside \textsc{QDecBench}.}
The bottom-up path needs no sketch, so it also applies to algorithms that \sys was not designed for. We test it on nine such families (Simon, Boolean hidden shift, swap test, HHL, two QUARK finance circuits, quantum volume, phase-oracle Grover, and quantum counting), 168 inputs compiled and checked like \textsc{QDecBench}. Checked structure explains 60\% of their gates, as layers, loops, multi-controlled gates, and library calls such as the QFTs inside HHL and counting, and the programs are 3.7$\times$ shorter than the input. Trivial-TM explains almost none of these circuits (Figure~\ref{fig:conditions}f). Naming the algorithm itself takes a sketch: without one, 26\% of the inputs are fully decompiled.

\paragraph{Extensibility.}
Adding a sketch is a local change that leaves the checker as it is. Classical decompilers grow the same way: IDA Pro, for example, recognizes library functions from signature files, and users add signatures for new libraries~\cite{flirt}. We added sketches for two of the nine families, Simon's algorithm and Boolean hidden shift, in 137 lines of Python. Over the nine families, they nearly double FD, from $0.256$ to $0.500$, and raise SC on both families (Figure~\ref{fig:conditions}f); the results on \textsc{QDecBench} are unchanged.

\subsection{RQ3: Contribution of Techniques}
\label{sec:rq3}
\begin{figure}[t]
\centering
\includegraphics[width=\textwidth]{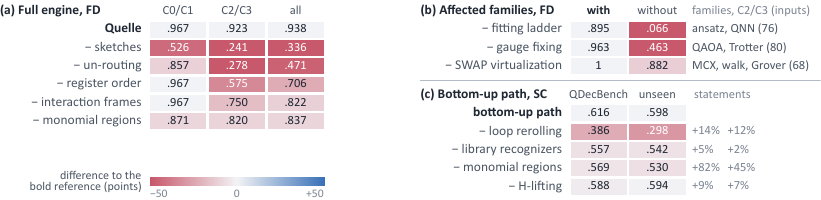}
\caption{Ablation study.}
\label{fig:ablation}
\end{figure}

\paragraph{Ablation (Figure~\ref{fig:ablation}).}
We remove one technique at a time and measure its effect on the inputs it acts on; as in Figure~\ref{fig:conditions}, a tile's colour shows its difference to the bold reference of its group.
\textbf{\textit{Full engine (a).}} Sketches carry most of the structure: without them, FD falls by 64\%. They read their parameters from the logical circuit, so they depend on un-routing. On the routed inputs (C2/C3), FD falls by 70\% without \texttt{SWAP} extraction and by 38\% without register-order recovery. The interaction frames matter only for native gates other than \texttt{cx}: without them, FD on the C2/C3 inputs compiled for \texttt{cz} and \texttt{ecr} falls by 35\% and 20\%. Removing monomial regions, which also disables H-lifting and semantic collapse, lowers FD by 11\%.
\textbf{\textit{Family-specific techniques (b).}} Gauge fixing, \texttt{SWAP} virtualization, and the fitting ladder act mainly on the routed and optimized inputs of a few families; at C0/C1, only the fitting ladder changes any output. On their families at C2/C3, the fitting ladder matters most: without it, FD on ans\"atze and QNNs falls by 93\%. Removing any one of them makes these programs 2.5--3.6$\times$ longer.
\textbf{\textit{Bottom-up path (c).}} With sketches disabled, each technique of this path contributes its own kind of structure. Rerolling produces the loops; without it, SC falls by 37\% on \textsc{QDecBench} and by 50\% on the unseen families. Library recognizers produce the QFT and state-preparation calls, and monomial regions the multi-controlled gates; without monomial regions, the programs become 1.5--1.8$\times$ as long. With sketches enabled, removing H-lifting or rerolling changes almost no output, because a sketch also explains nearly every input they restructure.

The consistency vote for cost-ambiguous \texttt{SWAP} blocks (\S\ref{sec:unroute}) changes one output on \textsc{QDecBench}. Over the 15 routed conditions of RQ2, it changes 15, all of them Trotter programs: with the vote, the Trotter sketch recovers the Hamiltonian and the step count; without it, the output is a single loop over two-qubit terms on the physical pairs left by routing, and the step count is lost.

\subsection{RQ4: Scalability and Real-World Assembly}
\label{sec:rq4}
\begin{figure}[t]
\begin{minipage}[b]{0.49\textwidth}
\centering
\includegraphics[width=\linewidth]{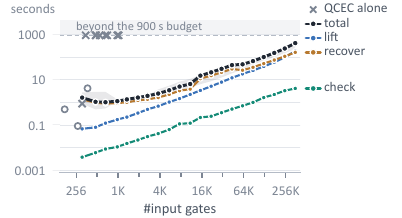}
\caption{Time per stage on random 12-qubit circuits.}
\label{fig:scale}
\end{minipage}\hfill
\begin{minipage}[b]{0.49\textwidth}
\centering
\includegraphics[width=\linewidth]{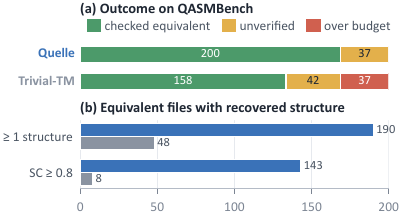}
\caption{Results on QASMBench.}
\label{fig:qasmbench}
\end{minipage}
\end{figure}

\paragraph{Scaling.}
We decompile random 12-qubit circuits compiled at C1, at 21 sizes up to 392K gates with three seeds each. The total time grows nearly linearly with the size of the input (Figure~\ref{fig:scale}), from about 6\,s at 10K gates to about 7 minutes at the largest size. Most of it is spent on lifting and recovery; the check remains a small part. A monolithic check does not scale in the same way. We tried MQT QCEC on 12 of the input/output pairs, none larger than 1{,}045 gates: it certifies 3 of them within 15 minutes, while \sys decompiles and checks each of the 12 in under 3\,s. On large adders, QCEC also reported two equivalent circuits as not equivalent (\S\ref{sec:verify}).

\paragraph{QASMBench.}
We also run \sys and Trivial-TM on the files of a pinned QASMBench~\cite{qasmbench} archive, excluding 15 of its 252 files that are not valid OpenQASM~2.0. The remaining 237 files reach 433 qubits and 5.8M gates, far beyond the benchmark inputs, so we relax the time budget to one hour per file. There is no ground truth, so we report the outcome of each run and the ground-truth-free structure measures (Figure~\ref{fig:qasmbench}).

\sys finishes every file within the hour, with a median time below one second; eight files take longer than five minutes. On the largest of them, two 24-qubit VQE circuits with more than 2M gates and two transpiled multipliers, rerolling uses up its share of the budget (\S\ref{sec:impl}), and most gates stay unrolled. Almost all outputs checked to be equivalent (190 of 200) contain recovered structure, and 143 reach SC $\ge 0.8$. Trivial-TM's output stays at the gate level and must be checked monolithically, mostly with MQT QCEC; on 37 files this check does not finish within the hour. Only 48 of its checked outputs contain any structure, and the 8 that reach SC $\ge 0.8$ are hand-written files of at most 22 gates.

\begin{figure}[t]
\begin{minipage}[t]{0.485\textwidth}
\begin{lstlisting}[style=py,basicstyle=\ttfamily\scriptsize]
def build_circuit() -> QuantumCircuit:
    qc = QuantumCircuit(5, 4, name='qpe')
    counting, target = [3, 2, 1, 0], 4
    qc.h(counting)
    phi = -3*pi/4
    for k, q in enumerate(counting):
        qc.cp(phi * 2**k, q, target)
        qc.p(3*pi/8 * 2**k, q)
    qc.compose(synth_qft_full(4, do_swaps=False,
                              inverse=True),
               [0, 1, 2, 3], inplace=True)
    qc.measure([0, 1, 2, 3], [0, 1, 2, 3])
    return qc
\end{lstlisting}
\centerline{\small (a) \texttt{pea\_n5\_transpiled} (108 gates)}
\vspace{4pt}
\begin{lstlisting}[style=py,basicstyle=\ttfamily\scriptsize]
def build_circuit() -> QuantumCircuit:
    qc = QuantumCircuit(11, 4, name='circuit')
    qc.h(range(0, 8))
    ...  # 16 statements, three of them ccx
    qc.mcp(pi, [1, 2, 4], 8)
    qc.mcp(pi, [0, 2, 3], 6)
    qc.cz(1, 2)
    ...  # 112 statements and the measurements
\end{lstlisting}
\centerline{\small (c) \texttt{sat\_n11\_transpiled} (731 gates)}
\end{minipage}\hfill
\begin{minipage}[t]{0.485\textwidth}
\begin{lstlisting}[style=py,basicstyle=\ttfamily\scriptsize]
def build_circuit() -> QuantumCircuit:
    qc = QuantumCircuit(7, 7, name='circuit')
    qc.ry(1.9412719087294656, 0)
    qc.h(range(1, 6))
    qc.s(6)
    for i in range(1, 6):
        qc.cp((376519*pi/518443) * 2**(i - 1),
              0, i)
    for i in range(5):
        qc.p(THETA0[i], i + 1)
    qc.h(6)
    qc.compose(synth_qft_full(3, do_swaps=False,
                              inverse=True),
               [5, 4, 3], inplace=True)
    ...  # 47 statements: rest of the inverse QFT,
    ...  # ancilla rotation (multi-qubit Z rotations)
    qc.compose(synth_qft_full(5, do_swaps=False),
               [5, 4, 3, 2, 1], inplace=True)
    for i in range(4):
        qc.cp(-(376519*pi/518443) * 2**i,
              0, i + 1)
        qc.p(THETA5[i], i + 1)
        qc.h(i + 1)
    ...  # 5 gates and the measurements
\end{lstlisting}
\centerline{\small (b) \texttt{hhl\_n7\_transpiled} (990 gates)}
\end{minipage}
\caption{Programs recovered from three QASMBench files.}
\label{fig:cases}
\end{figure}

\emph{Case studies.} Figure~\ref{fig:cases} shows three recovered programs.
(a)~\texttt{pea\_n5\_transpiled} is the textbook phase estimation: a Hadamard
layer, a loop of controlled phases whose angle \sys solves as $\varphi=-3\pi/4$,
and an inverse QFT. (b)~\texttt{hhl\_n7\_transpiled} (SC $0.79$) exposes most of
the HHL skeleton: the same phase-estimation loop and inverse QFT, the ancilla
rotation (omitted), and the uncomputation, which repeats the loop with negated
angles. (c)~\texttt{sat\_n11\_transpiled} (SC $0.54$) has an irregular oracle:
\sys recovers its Toffoli and multi-controlled phase gates without sketchs.

\subsection{RQ5: Downstream Applications}
\label{sec:rq5}

\paragraph{Comprehension.}
We use GPT-6.1-Sol, the stronger model in RQ1, as a proxy reader. Given either
the raw assembly or \sys's output, and without being told the algorithm, it is
asked (L1)~which algorithm an input implements and (L2)~how many times its main
repeated block is applied, such as the layers of a QAOA circuit. L1 is scored on
all 816 inputs (for random circuits the right answer is ``no known algorithm''),
and L2 on the 348 inputs whose ground truth records a repetition count.

\begin{figure}[t]
\begin{minipage}[t]{0.42\textwidth}
\vspace{0pt}
\includegraphics[width=\linewidth]{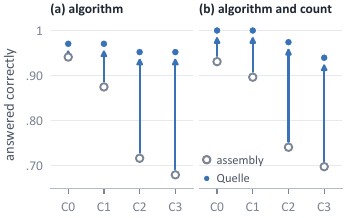}
\caption{LLM comprehension (\mbox{GPT-6.1-Sol}).}
\label{fig:ident}
\end{minipage}\hfill
\begin{minipage}[t]{0.56\textwidth}
\vspace{0pt}
\includegraphics[width=\linewidth]{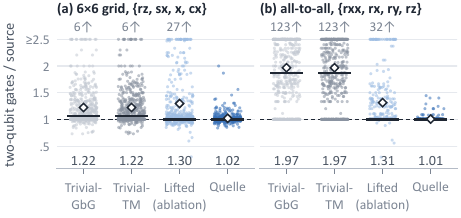}
\caption{Retargeting routed assembly.}
\label{fig:retarget}
\end{minipage}
\end{figure}

From the raw assembly, the model identifies the algorithm less often the further
compilation moves from the source, from 94\% of the inputs at C0 to 68\% at C3;
from \sys's output, it identifies 95--97\% at every level
(Figure~\ref{fig:ident}a). Answering both questions shows the same pattern
(Figure~\ref{fig:ident}b). The gap is largest where compilation hides the
structure: from the raw assembly, the model recognizes amplitude estimation in
only 3 of 24 inputs and gets just over half of the Grover and quantum-walk
counts right, while from \sys's output it answers all of them correctly. Reading
\sys's output also takes 7$\times$ fewer tokens.

\paragraph{Retargeting.}
Assembly compiled for one device is often run on another. We recompile each
\textsc{QDecBench} input for two targets, a $6\times 6$ grid with
$\{\texttt{rz},\texttt{sx},\texttt{x},\texttt{cx}\}$ and an all-to-all
trapped-ion device with $\{\textsf{rxx},\texttt{rx},\texttt{ry},\texttt{rz}\}$,
with Qiskit at optimization level~3 and three seeds. We start from (i)~the
assembly itself (the Trivial-GbG program), (ii)~the Trivial-TM program,
(iii)~\sys's lifted gate-level program without recovered structure (an
ablation), and (iv)~\sys's recovered program, and compare with compiling (v)~the
source program. We compile without assuming that idle qubits start in
$|0\rangle$, so that a result can be checked against the program it came from.
We report the mean over the seeds and check every one of the 24{,}480 results:
all agree with their programs, 23{,}761 certified by MQT QCEC and 719, where
QCEC is inconclusive, by simulation on random input states.
Figure~\ref{fig:retarget} shows the two-qubit gate count of each recompiled
routed input relative to the recompiled source, averaged over the three seeds.
Recompiling the assembly carries the first device's routing into the second: it
needs $1.22\times$ the two-qubit gates of the source on the grid and
$1.97\times$ on the ion device, which needs no routing at all. Trivial-TM gives
the same counts, because its templates replace contiguous gate blocks and the
compiler receives the same circuit. From \sys's program, the count is within 2\%
of the source. The lifted program alone needs about $1.3\times$ the gates,
because its canonical multi-controlled and phase gates (\S\ref{sec:canon}) cost
more to synthesize than the recovered library operations and loops. For a
7-qubit QAOA program, for example, the assembly retargets to the ion device with
172 two-qubit gates and \sys's program with 45, as many as the source. On
unrouted inputs (C0/C1) the differences are smaller.

\section{Discussion and Limitations}
\label{sec:discussion}

\paragraph{Potential Applications.}
Besides the comprehension and retargeting evaluated in \S\ref{sec:rq5}, a program checked for equivalence supports \emph{auditing} (does published assembly implement the claimed algorithm?), \emph{regression testing of compilers} (decompile the output of two compiler versions and compare the recovered programs), and \emph{reuse} (extract a recovered QFT or adder as a subroutine).

\paragraph{Guarantees.}
The guarantee is about behavior, in the sense of translation
validation~\cite{pnueli}: a program reported as \texttt{equivalent} does exactly
what the input assembly does on an ideal device, as established by a check that
is independent of the recovery (\S\ref{sec:invariant}). The recovered structure
describes the input faithfully, but it is not necessarily the structure the
programmer wrote. Classical decompilers face the same problem: once a compiler
emits the same instructions for two different source expressions, nothing in the
binary tells them apart. In quantum programs this happens when two algorithms
compute the same operation, as two textbook adder constructions do at small
sizes; \sys then reports the one whose check passes first. Likewise, code that
the compiler deletes because it does not affect the result, such as the oracle
of a Deutsch--Jozsa program whose oracle is constant, cannot be recovered by any
decompiler.

\paragraph{Limitations.}
(1)~\emph{Dynamic circuits.} \sys emits programs with mid-circuit measurement, reset, or classical control, but does not check them; every QASMBench file reported \texttt{unverified} is of this kind.
(2)~\emph{Algorithm names need a sketch.} Without a sketch, \sys still recovers loops, library calls, and multi-controlled gates, but does not name the algorithm. Naming a new algorithm takes a small solver; the checker needs no change (\S\ref{sec:rq2}).
(3)~\emph{Obfuscation.} We assume a benign compiler. Deliberately obfuscated assembly, for example with inserted gate pairs that cancel, is outside our scope. Soundness does not depend on this assumption, but the amount of recovered structure does.

\paragraph{Threats to validity.}
The benchmark is compiled with Qiskit; the further compilation settings of RQ2,
including a second compiler (tket), and the hand-written QASMBench files reduce
this dependence. Equivalence is checked up to a numerical tolerance of
$10^{-6}$. LLM results vary between runs and model versions; we report the
versions and average over three runs.

\section{Related Work}
\label{sec:related}

\paragraph{Classical and learned decompilation.}
Decompilers recover types~\cite{tie} and control structure~\cite{phoenix} from
binaries, and their output is judged by readability~\cite{cifuentes};
learning-based decompilers, including LLM-based ones, translate assembly to
source~\cite{llm4decompile,decllm,jiang2025nova}, recover variable names, types, and data
structures~\cite{chen2022dirty,xie2024resym}, or polish decompiler output~\cite{hu2024degpt}.
The correctness of decompiler output is an active research topic: studies measure its semantic
consistency and fidelity~\cite{cao2024evaluating,dramko2024taxonomy}, examine what recompilation and
tests establish~\cite{liu2026recompile}, and validate neural decompilers by program
equivalence~\cite{dramko2026faultless}.
BTD~\cite{btd} decompiles deep-learning
executables to model specifications by identifying operators and recovering
their parameters. For quantum programs, DeQompile~\cite{deqompile} evolves a
program that generates the circuits of one algorithm by genetic programming,
scored by syntactic similarity, and model-driven reverse engineering abstracts
quantum source programs into KDM models~\cite{kdm}. \sys decompiles a single
compiled program and, unlike these decompilers, checks its output against the input
under exact equivalence.

\paragraph{Quantum software engineering.}
Qiskit~\cite{qiskit} and tket~\cite{tket} lower, route~\cite{sabre}, and
optimize circuits; ZX-calculus rewriting~\cite{pyzx} and phase-polynomial
optimizers~\cite{amy,nam} reduce gate counts, template matching finds
subcircuits modulo gate commutation~\cite{iten}, and inverse transpilation
infers which optimization passes produced a circuit~\cite{inversetranspile}.
Verified compilers and optimizers prove such passes
correct~\cite{voqc,certiq,giallar}, and equivalence checkers decide whether two
circuits agree through decision diagrams or ZX reasoning~\cite{qcec}. Around
this tool chain, quantum software engineering has built an ecosystem of testing,
benchmarking~\cite{qasmbench,mqtbench}, and analysis: differential testing of
software stacks~\cite{wang2021qdiff}, fuzzing of
simulators~\cite{klimis2025simulators}, test generation, mutation analysis, and
property-based testing for quantum
programs~\cite{wang2021quito,mendiluze2021muskit,ye2023quratest,honarvar2020pbt,pontolillo2024deltadebug},
static analysis~\cite{paltenghi2024lintq}, code-smell studies, including the
effect of
transpilation~\cite{chen2023smelly,destefano2024transpilation,destefano2022howfar},
and LLM-based repair~\cite{yoshida2026mutation}.
IterTestQ~\cite{paltenghi2026itertestq} is closest in setting: it works directly
on OpenQASM assembly, importing it into a platform, transforming it, exporting
it again, and comparing the results with an equivalence oracle. \sys works in
the inverse direction, which uses phase polynomials and equivalence checking as
components.

\paragraph{Program synthesis and verified lifting.}
Sketch-based synthesis~\cite{sketch} fills holes in a template by constraint solving. Verified lifting synthesizes a high-level summary of low-level code and proves it equivalent, for stencils~\cite{stng}, MapReduce programs~\cite{casper}, and tensor operations~\cite{tenspiler}, within a general framework~\cite{metalift}, and with summaries proposed by LLMs~\cite{llmlift}.
\sys applies the same propose-and-check principle to quantum assembly, with readability as the goal and sketch parameters solved from circuit invariants; its loop rerolling builds on anti-unification~\cite{plotkin}.

\section{Conclusion}
\label{sec:conclusion}
We presented \sys, a decompiler that turns quantum assembly back into structured programs and checks each of them for equivalence with its input. Because compilation preserves the semantics of a program while changing its syntax, compiler artifacts can be removed by exact local rewrites, algorithm parameters can be solved from invariants of the lifted circuit, and every recovery hypothesis can be checked with an exact equivalence oracle. On 816 benchmark programs and 237 real-world assembly files, \sys emitted no inequivalent program; it fully decompiles 94\% of the benchmark inputs, 1.7$\times$ as many as the best LLM-based decompiler, at no model cost. We hope decompilation will support auditing, retargeting, and comprehension of the quantum assembly the community already exchanges.

\bibliographystyle{ACM-Reference-Format}
\bibliography{refs}

\newpage
\appendix
\section{Gate Reference}
\label{app:gates}
Table~\ref{tab:gates} lists the gates that appear in this paper, grouped by how they act on bit strings, with the circuit symbols used in the figures.

\begin{table}[h]
\caption{Gates used in this paper.}
\label{tab:gates}
\footnotesize\renewcommand{\arraystretch}{1.45}\setlength{\tabcolsep}{4pt}
\begin{tabular}{@{}l l c l p{3.7cm}@{}}
\toprule
Kind & Gate & Symbol & Action on basis states & Meaning \\
\midrule
\multirow{4}{*}{Permutation}
 & \texttt{x} (NOT) & \qbox{X} & $|b\rangle\mapsto|\lnot b\rangle$ & bit flip \\
 & \texttt{cx}$(a,b)$ (CNOT) & \qcx & $|a,b\rangle\mapsto|a,\;a\oplus b\rangle$ & \texttt{b \textasciicircum= a} \\
 & \texttt{ccx}$(a,b,c)$ (Toffoli) & \qccx & $|a,b,c\rangle\mapsto|a,b,\;c\oplus ab\rangle$ & \texttt{c \textasciicircum= a \& b}; with more controls, \texttt{mcx} \\
 & \texttt{swap}$(a,b)$ & \qswap & $|a,b\rangle\mapsto|b,a\rangle$ & exchange two qubits; costs three \texttt{cx} \\
\midrule
\multirow{3}{*}{Phase}
 & \texttt{p}$(\lambda)$, \texttt{rz}$(\lambda)$; \texttt{z}, \texttt{s}, \texttt{t} & \qbox{P} & $|b\rangle\mapsto e^{i\lambda b}|b\rangle$ & phase when the bit is 1; $\lambda=\pi,\frac{\pi}{2},\frac{\pi}{4}$ for \texttt{z}, \texttt{s}, \texttt{t} \\
 & \texttt{cp}$(\lambda)$, \texttt{cz} & \qcz & $|a,b\rangle\mapsto e^{i\lambda ab}|a,b\rangle$ & phase when both bits are 1; \texttt{cz} is \texttt{cp}$(\pi)$ \\
 & \texttt{rzz}$(\lambda)$ & \qbbox{ZZ} & $|a,b\rangle\mapsto e^{i\lambda(a\oplus b)}|a,b\rangle$ & phase by parity \\
\midrule
\multirow{2}{*}{Mixing}
 & \texttt{h} (Hadamard) & \qbox{H} & $|b\rangle\mapsto\tfrac{1}{\sqrt2}\big(|0\rangle+(-1)^b|1\rangle\big)$ & uniform superposition from $|0\rangle$; swaps the roles of \texttt{x} and \texttt{z} \\
 & \texttt{rx}$(\theta)$, \texttt{ry}$(\theta)$, \texttt{sx} & \qbox{R} & $e^{-i\theta X/2}$, $e^{-i\theta Y/2}$; $\texttt{sx}=\sqrt{\texttt{x}}$ & rotations that mix $|0\rangle$ and $|1\rangle$ \\
\bottomrule
\end{tabular}
\\[2pt]{\footnotesize \texttt{rz} and \texttt{rzz} are listed up to a global phase: $\texttt{rz}(\lambda)=e^{-i\lambda Z/2}$ and $\texttt{rzz}(\lambda)=e^{-i\lambda Z\otimes Z/2}$.}
\end{table}

\section{Algorithm Sketches}
\label{app:sketches}
Table~\ref{tab:sketches} lists the sketches of \S\ref{sec:sketch} with the parameters that each solver reads from the lifted circuit.

\begin{table}[h]
\caption{Algorithm sketches.}
\label{tab:sketches}
\small
\begin{tabular}{@{}p{2.4cm}p{\dimexpr\textwidth-2.4cm-2\tabcolsep\relax}@{}}
\toprule
Sketch & Parameters and solver \\
\midrule
Graph/GHZ/W & CZ edge set; CX fan-out tree; cascade of controlled rotation angles\\
BV / DJ & kickback targets from the oracle's CX set (constant oracles are erased by the compiler, \S\ref{sec:eval})\\
MCX & control state $s$: $U H_A$, with $H_A$ the Hadamard layer on the register $A$, is a permutation that flips the target exactly when $x_C=s$\\
Grover & marked state and flag qubit; iteration count $k\in\{1,2,3,k_0\pm1\}$ by checking each candidate\\
QPE / AE & eigenphase and state preparation from the block-diagonal structure of $U$\\
QFT state & semantic factoring $U=\mathrm{QFT}\cdot X^s$\\
Quantum walk & coin, node order, depth\\
Arithmetic & register sizes; adder/multiplier variant from the library\\
QAOA / Trotter & Pauli-rotation normal form (\S\ref{sec:pauli})\\
Ans\"atze / QNN & template instantiation modulo gauge (\S\ref{sec:fit})\\
\bottomrule
\end{tabular}
\end{table}

\section{QDecBench Families}
\label{app:bench}
Table~\ref{tab:families} lists the 22 algorithm families of \textsc{QDecBench} in their three tiers, and the random circuits that serve as a negative control (NC). Each source program is compiled at the four levels C0--C3 (\S\ref{sec:setup}); the inputs column counts the resulting assembly programs. ``(count)'' marks a ground-truth structure that also records its number of repetitions.

\paragraph{Negative control.} The random circuits are generated by Qiskit's \texttt{random\_circuit} with depth~6 and at most two qubits per gate: each layer applies gates drawn at random from Qiskit's standard gate library, with random angles, to randomly chosen qubits, and all qubits are measured at the end. We use two instances (5 qubits with seed~1, 22 gates; 7 qubits with seed~2, 25 gates), compiled like every other program, which gives 12 inputs. They implement no algorithm, so the expected output is an equivalent program that claims none; a sketch that matched one of them would be a false recognition. An enlarged negative control outside the 816 inputs adds 104 random circuits generated the same way, 13 seeds (10--22) for each size from 5 to 12 qubits, compiled at C0--C3 like \textsc{QDecBench} (624 inputs, each checked against its source).

\begin{table}[h]
\caption{QDecBench families.}
\label{tab:families}
\footnotesize\renewcommand{\arraystretch}{1.1}\setlength{\tabcolsep}{4pt}
\begin{tabular}{@{}l l l c r p{3.2cm} p{2.9cm}@{}}
\toprule
Tier & Family & Source & Qubits & Inputs & Varied across instances & Ground-truth structure \\
\midrule
T1 & GHZ state & MQT Bench & 5--12 & 30 & size & GHZ preparation \\
 & W state & MQT Bench & 5--8 & 18 & size & W-state preparation \\
 & Graph state & MQT Bench & 6--10 & 42 & random graph & graph-state preparation \\
 & QFT & Qiskit & 4--8 & 108 & QFT, inverse, without swaps, approximate; input state & QFT or inverse QFT \\
 & MCX & Qiskit & 4--7 & 60 & control state & multi-controlled X \\
 & Draper adder & MQT Bench & 6--8 & 12 & size & adder; QFT; inverse QFT \\
 & Cuccaro adder & MQT Bench & 6--8 & 12 & size & ripple-carry adder \\
 & VBE adder & MQT Bench & 7--10 & 12 & size & ripple-carry adder \\
\midrule
T2 & Bernstein--Vazirani & MQT Bench & 6--12 & 54 & hidden string & BV \\
 & Deutsch--Jozsa & MQT Bench & 5--11 & 54 & balanced or constant oracle & DJ \\
 & QPE & MQT Bench & 5--9 & 60 & exact or inexact phase & QPE; controlled powers (count); inverse QFT \\
 & Amplitude estimation & MQT Bench & 4--7 & 24 & success probability & AE; controlled powers (count); inverse QFT \\
 & Grover & MQT Bench & 4--7 & 24 & size & Grover; iterations (count) \\
 & QAOA & MQT Bench & 5--8 & 60 & graph, 2--3 layers, angles & QAOA; layers (count) \\
 & Variational ansatz & MQT Bench & 4--7 & 102 & 3 ans\"atze, 4 entanglements, 2--3 repetitions, angles & ansatz; repetitions (count) \\
 & QNN & MQT Bench & 4--6 & 12 & parameters & feature map; ansatz \\
 & Trotterized evolution & Qiskit & 4--7 & 60 & Ising or Heisenberg chain, couplings, time, 2--4 steps & Trotter; steps (count) \\
 & Entangled QFT & MQT Bench & 5--7 & 12 & size & GHZ preparation; QFT \\
\midrule
T3 & Quantum walk & MQT Bench & 4--5 & 18 & 2--3 steps & walk steps (count) \\
 & Modular adder & MQT Bench & 6--8 & 12 & size & modular adder \\
 & Half adder & MQT Bench & 5--7 & 12 & size & half adder \\
 & Multiplier & MQT Bench & 4 & 6 & --- & QFT-based multiplier \\
\midrule
NC & Random circuit & Qiskit & 5--7 & 12 & seed & none \\
\bottomrule
\end{tabular}
\end{table}

\begin{table}[h]
\caption{Compilation levels.}
\label{tab:levels}
\footnotesize\setlength{\tabcolsep}{3pt}
\begin{tabular}{@{}l l l l c c@{}}
\toprule
Level & Gate set & Coupling map & Layout and routing & Opt. & Seeds \\
\midrule
C0 & $u_3$, \texttt{cx} & none & none & 0 & 1 \\
C1 & \texttt{rz}, \texttt{sx}, \texttt{x}, \texttt{cx} & none & none & 1 & 1 \\
C2 & \texttt{rz}, \texttt{sx}, \texttt{x}, and \texttt{cx}, \texttt{ecr}, or \texttt{cz} & line, grid, heavy-hex & VF2/SABRE, SABRE & 1--2 & 2 \\
C3 & \texttt{rz}, \texttt{sx}, \texttt{x}, and \texttt{cx}, \texttt{ecr}, or \texttt{cz} & line, grid, heavy-hex & VF2/SABRE, SABRE & 3 & 2 \\
\bottomrule
\end{tabular}
\end{table}

\section{Compilation Levels}
\label{app:levels}
Table~\ref{tab:levels} lists the Qiskit transpiler settings of the four compilation levels. C0 only translates the program to $\{u_3,\texttt{cx}\}$, so its assembly stays close to the source. C1 targets the gate set of earlier IBM devices, $\{\texttt{rz},\texttt{sx},\texttt{x},\texttt{cx}\}$, and merges and cancels gates. C2 and C3 also map the program onto a device: the coupling map is a connected subgraph with exactly as many physical qubits as the program has qubits, cut from a line, a grid, or a heavy-hex lattice, the three chosen in rotation over programs and seeds. The native two-qubit gate rotates in the same way over \texttt{cx}, \texttt{ecr} (IBM Eagle), and \texttt{cz} (IBM Heron), shifted by one at C3, so that across C2 and C3 every coupling map meets two of the three gates. As for a real device, the transpiler chooses the initial layout itself (VF2Layout, which places the program without \texttt{SWAP}s when its interactions fit the map, else SabreLayout), and SABRE routing inserts \texttt{SWAP}s for the remaining interactions. C2 optimizes at level~1 or~2, and C3 at level~3, which resynthesizes two-qubit blocks and therefore changes the gate-level structure most. C0 and C1 involve no randomness and are compiled once per program; C2 and C3 are compiled with two transpiler seeds, and C3 offsets its seed so that it does not repeat the layout and routing of C2. A program thus yields six assembly programs.
Every compiled program is checked against its source: its unitary, with the initial layout and the final routing permutation applied, must equal the source unitary up to global phase.

\section{Structured Fraction}
\label{app:sf}
SF is computed from the output program alone, so it applies to every tool in the same way. We execute the program under a line tracer and record, for each gate of the circuit it returns, the source line and call stack that appended it. A gate counts as \emph{structured} if it was appended by a statement inside a loop that ran at least twice, inside a user-defined function that was called at least twice, by a single statement that applies the same gate to two or more qubits, or by a library-level operation (three or more qubits, or a composite library gate). Each gate is weighted by its size after decomposition to $\{u,\texttt{cx}\}$, so that a QFT on five qubits weighs more than one rotation, and SF is the structured fraction of the total weight. Two patterns never count as structure, because they let a program look structured while it only replays a transcription: loops that look up gate names or arguments in a data table (\texttt{getattr}, \texttt{eval}), and gates given as explicit matrices (\texttt{UnitaryGate}, \texttt{initialize}).

\section{The Trivial-TM Baseline}
\label{app:tm}
Trivial-TM recovers structure by matching templates directly against the compiled circuit, with none of \sys's lifting steps (no un-routing, no canonicalization), so it shows what structure recovery achieves when compilation artifacts are left in place. It tries the following matchers in order and keeps the first that succeeds.
(1)~\emph{Whole circuit.} For circuits of up to 12 qubits and 1{,}500 gates, the unitary part of the circuit, optionally after up to three leading single-qubit gates, is compared with templates of matching width: GHZ, uniform-superposition, and basis-state preparation; the Grover diffusion operator; Toffoli, relative-phase Toffoli, and Fredkin gates; multi-controlled X; \textsf{rzz} and \textsf{rxx} rotations with the angles that occur in the circuit; and the QFT and its variants (inverse, without the final swaps, approximation degree 0--2).
(2)~\emph{GHZ region.} A Hadamard followed by a chain or fan-out of CNOTs anywhere in the circuit.
(3)~\emph{Repetition.} A contiguous block of gates that repeats exactly is written as a loop.
(4)~\emph{Normalized subroutine.} For circuits of up to 10 qubits and 256 gates, one subroutine is matched after a bounded normalization of the gate sequence.
Every match is accepted only if the unitaries agree exactly (up to global phase); the gates outside the match are transcribed one by one, and the whole program is checked against the input like \sys's output.

\section{DeQompile}
\label{app:deqompile}
DeQompile~\cite{deqompile} is presented by its authors as a quantum decompiler. We ran its released code (the latest commit, \texttt{0c156e0}) as shipped and scored every output with the procedure of the LLM baselines (\S\ref{sec:setup}).

\paragraph{What it computes.}
Given the circuits of one algorithm at the sizes $n=2,\ldots,20$, DeQompile evolves a Python function $P(n)$ that builds a circuit on $n$ qubits: a program of at most ten statements, with loops nested at most twice, over the gate set $\{\texttt{h},\texttt{x},\texttt{cx}\}$. A candidate is scored by comparing the OpenQASM text it prints with the target file line by line, averaged over the sizes and reduced by a length penalty; no candidate is checked for equivalence. A run evolves 100 generations of 50 candidates.

\paragraph{Why it does not apply.}
Three properties of this design conflict with the decompilation of compiled assembly.
(1)~It generalizes over a family of circuits of different sizes, whereas a decompiler receives one program, and an assembly file comes without compilations of the same algorithm at other sizes.
(2)~Its objective is textual similarity, so a candidate can score well without being equivalent to the target, and the tool does not report whether it is.
(3)~Compilation rewrites the text that this objective compares: routing, basis translation, and gate merging change the gate sequence and preserve the unitary (\S\ref{sec:compile}), so the objective is maximized by a program that prints the compiled gate sequence itself, not by the source program.
In addition, the default gate set cannot express the native gates of the inputs.

\paragraph{Setup.}
We changed none of DeQompile's settings: hyperparameters, gate set, fitness function, random seed, and time limit are as shipped. Two adaptations were necessary to run it. First, DeQompile reads a target file for every size from 2 to 20 and stops when one is missing; we restricted this loop to the sizes present in a group (one or two). Second, its programs build circuits without classical bits, so each scored program calls $P(n)$ at the input's $n$ and appends the input's measurements. A group is one family, variant, and level, and at C2/C3 also one transpiler seed. DeQompile runs in its own environment (Qiskit 1.3.2), and its programs are scored in ours.

\paragraph{Results.}
On the 360 test inputs (210 runs, each completing its 100 generations in 37--77\,s), no returned program is equivalent to its input, so FD(SF) is 0. At the input's size, 286 programs raise an error (a two-qubit gate whose two qubit expressions evaluate to the same qubit), 70 build no gate, and 4 build gates that do not implement the input. The programs also score poorly on DeQompile's own objective: their textual similarity to the target is at most 0.008, where 1 means identical text. Every input contains gates outside the default gate set (\texttt{u3} at C0; \texttt{rz} and \texttt{sx} at C1--C3, and \texttt{ecr} or \texttt{cz} at C2/C3 for most inputs), and these gates make up 85\% of the inputs' gate lines. The line-by-line comparison starts at the fourth line of each file, which for 340 of the 360 inputs is a register declaration that DeQompile never prints; in 144 of the 210 runs, no generation improved on the best candidate of the random initial population.

\end{document}